\documentclass[
    prb,
  onecolumn,
  superscriptaddress,
  floatfix
]{revtex4-2}

\usepackage{amsmath,amssymb}
\usepackage{graphicx}
\usepackage{subfig} 
\usepackage{xcolor}
\usepackage{soul}
\usepackage{ulem}
\usepackage{changes}

\def\a3913{$A_3$V$_9$Te$_{13}$}
\def\cs3913{Cs$_3$V$_9$Te$_{13}$}
\def\rb3913{Rb$_3$V$_9$Te$_{13}$}

\begin{document}

\title{Interlocked Vortex Order in the Frustrated $A_3$V$_9$Te$_{13}$ Lattice ($A$=Cs,Rb)}

\author{Zilong Li}
\affiliation{New Cornerstone Science Laboratory, Center for Correlated Matter, Zhejiang University, Hangzhou 310058, China}
\affiliation{School of Physics, Zhejiang University, Hangzhou 310058, China}

\author{H.-Q. Yuan}
\affiliation{New Cornerstone Science Laboratory, Center for Correlated Matter, Zhejiang University, Hangzhou 310058, China}
\affiliation{School of Physics, Zhejiang University, Hangzhou 310058, China}
\affiliation{Institute for Advanced Study in Physics, Zhejiang University, Hangzhou 310058, China}

\author{Xiaoqun Wang}
\affiliation{School of Physics, Zhejiang University, Hangzhou 310058, China}
\affiliation{Institute for Advanced Study in Physics, Zhejiang University, Hangzhou 310058, China}

\author{Chao Cao}
\email{ccao@zju.edu.cn}
\affiliation{New Cornerstone Science Laboratory, Center for Correlated Matter, Zhejiang University, Hangzhou 310058, China}
\affiliation{School of Physics, Zhejiang University, Hangzhou 310058, China}
\affiliation{Institute for Advanced Study in Physics, Zhejiang University, Hangzhou 310058, China}

\date{\today}

\begin{abstract}
Recently synthesized \a3913\ ($A$ = Cs, Rb) provides a unique platform for exploring the interplay between geometric frustration, magnetism, and correlated electronic states beyond conventional kagome systems. Based on first-principles calculations, we demonstrate that the frustrated V sublattice naturally stabilizes a robust vortex or anti-vortex antiferromagnetic ground state, whose total energy is substantially lower than all other configurations. In addition, the Heisenberg model analysis show strong constraints on the ground state magnetic configuration, such that fixing spin configuration of 1 triangle propagates and fixes all others. The vortex or anti-vortex phase hosts electronic structures reminiscent of kagome metals, including Dirac-like band crossings, van Hove singularities, and flat bands near the Fermi level. Owing to the noncollinear spin texture and quasi-two-dimensional magnetic character, the Néel temperature exhibits an extremely weak dependence on both the magnitude and direction of external magnetic fields, consistent with experimental observations. For Cs-based compound, the ground state band structure is metallic with 6 Weyl points in the BZ; while for the Rb-based compound, spin--orbit coupling opens an approximately 27-meV gap at ambient pressure due to the lifted C$_3$ symmetry. Our results establish \a3913\ as a promising platform for studying frustration-driven chiral magnetism and its interplay with topological electronic states.
\end{abstract}
\maketitle

\section{Introduction}
Kagome lattices have ushered in a new era of quantum materials, exhibiting a rich interplay among geometric frustration, electronic topology, and electron correlation effects\cite{kagome_1, mielkeFerromagneticGroundStates1991,ranProjectedWaveFunctionStudySpin2007, balentsSpinLiquidsFrustrated2010, yanSpinLiquidGroundState2011, yuChiralSuperconductingPhase2012, wangCompetingElectronicOrders2013, xuIntrinsicQuantumAnomalous2015, liRealizationFlatBand2018, yinTopologicalKagomeMagnets2022, wangQuantumStatesIntertwining2023,negiMagneticKagomeMaterials2025}. It can host Dirac cones, van Hove singularities (VHS), and flat-band electronic states, which promote strong correlation and topological phenomena\cite{yeMassiveDiracFermions2018, kangDiracFermionsFlat2020, huRichNatureVan2022, kangTwofoldVanHove2022, yangObservationFlatBand2023, wilsonAV3Sb5KagomeSuperconductors2024, yeHoppingFrustrationinducedFlat2024, disanteKagomeMetals2026, pengFlatBandsDistinct2026, wangSpinExcitationsFlat2025, wuFlatbandEnhancedAntiferromagnetic2025a}, as exemplified by $A$V$_3$Sb$_5$ ($A$=K, Rb, Cs)\cite{ortizNewKagomePrototype2019}, also known as the ``135" family of kagome compounds. A variety of intertwined electronic phases, including charge-density wave (CDW)\cite{tanChargeDensityWaves2021, mielkeTimereversalSymmetrybreakingCharge2022, zhengEmergentChargeOrder2022, wangSoftPhononChargeDensityWave2026}, electronic nematicity\cite{nieChargedensitywavedrivenElectronicNematicity2022, huNontrivialBandTopology2023, grandiTheoryNematicCharge2023, yangSuperconductivityNematicOrder2024}, and superconductivity\cite{ortizCs135_Z2Topological2020, chenRotonPairDensity2021, ortizSuperconductivityZ2Kagome2021, chenDoubleSuperconductingDome2021, duanNodelessSuperconductivityKagome2021, niAnisotropicSuperconductingProperties2021, yinSuperconductivityNormalStateProperties2021, neupertChargeOrderSuperconductivity2022, liuSuperconductivityPressureChromiumbased2024, liElectronCorrelationIncipient2025}, have been observed in these materials, establishing kagome systems as a versatile platform for exploring magnetism, topology, superconductivity, electron-phonon coupling, and correlated electron physics\cite{wangCompetingElectronicOrders2013, wangCompetitionChargedensitywaveSuperconductivity2021, zhaoCascadeCorrelatedElectron2021, luoElectronicNatureCharge2022, guoSwitchableChiralTransport2022, xuAltermagneticGroundState2025, xieElectronCorrelationsKagome2025, guoManybodyInterferenceKagome2025, liuUnravelingIntertwinedOrders2026, chengFriezeChargeStripes2026, huangControllingAltermagneticSpin2026, liExoticSurfaceStripe2025}.

\begin{figure*}[t]
    \includegraphics[width=\textwidth]{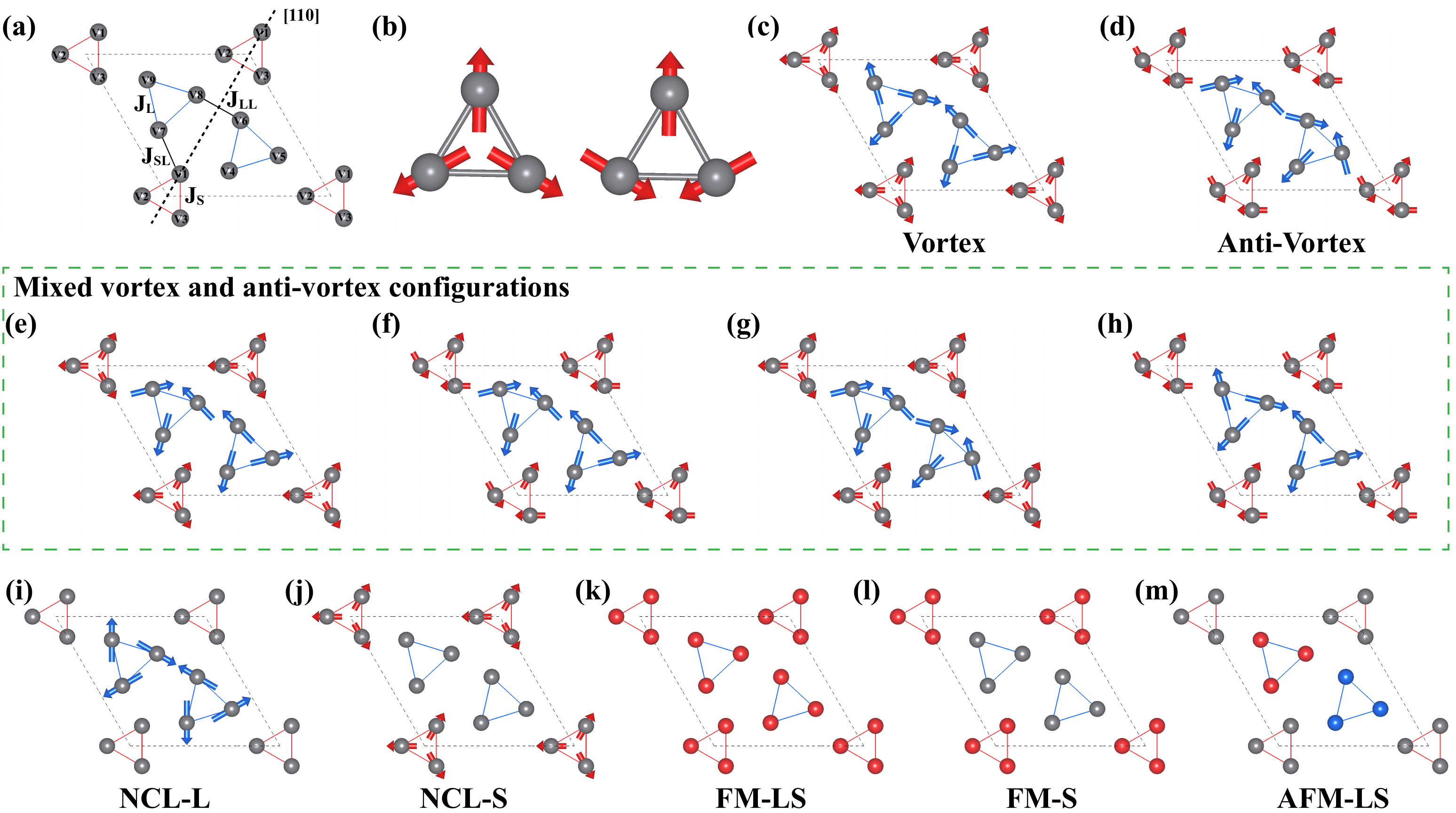}
    \caption{(a) Geometry of V-sublattice in \cs3913. The dominant nearest-neighbor exchange interactions and atom indices are labeled. (b) Schematic illustrations of the coplanar $120^\circ$ spin textures with opposite vector chiralities on an isolated triangle. When the triangle is traversed clockwise, the spin directions rotate clockwise by $120^\circ$ for the vortex configuration (left panel) and counterclockwise by $120^\circ$ for the anti-vortex configuration (right panel). (c-m) Magnetic configurations considered in this work. Noncollinear configurations include: (c) Uniform vortex, (d) uniform anti-vortex, (e)--(h) mixed vortex/anti-vortex configurations, (i) NCL-L, and (j) NCL-S.  Collinear configurations include: (k) FM-LS, (l) FM-S, and (m) AFM-LS. Red and blue arrows indicate the spin directions on the small- and large-triangle V sites, respectively. Red and blue spheres in (k)--(m) distinguish V sites with spin up and spin down.}
    \label{fig:fig1}
\end{figure*}

Recently, a ternary vanadium telluride compound \cs3913\ with a Reuleaux-triangle-like structure has been synthesized experimentally\cite{xiaoHeavyFermionBehaviorTunable2026, zhaoCs3913_NewVanadiumBased2026, liuCs3913CorrelatedElectron2026}. In contrast to the conventional kagome lattice composed of corner-sharing triangles, the V sublattice in Cs$_3$V$_9$Te$_{13}$ consists of two inequivalent V sites forming a two-dimensional network [Fig.~\ref{fig:fig1}(a)]. These V atoms form 2 large triangles and 1 small triangle in each unit cell, each retaining a local $C_3$ rotational symmetry about its center. The resulting structure constitutes a distinct frustrated lattice characterized by an intertwined arrangement of triangles, quadrilaterals, and pentagons. Transport, Hall, and magnetic measurements reveal that \cs3913\ undergoes a pronounced phase transition around 48~K\cite{xiaoHeavyFermionBehaviorTunable2026, zhaoCs3913_NewVanadiumBased2026, liuCs3913CorrelatedElectron2026}. No detectable structural distortion has been observed across the transition, indicating that its origin is likely electronic or magnetic in nature\cite{zhaoCs3913_NewVanadiumBased2026}. However, the transition temperature exhibits a remarkably weak dependence on external magnetic field. Quasi-two-dimensional electronic structure can be inferred from electrical resistivity measurements, where pronounced anisotropy of resistivities along the $c$-axis ($\rho_c$) and in-plane ($\rho_{ab}$) is observed. Replacing Cs with Rb leads to \rb3913, which adopts an AB-stacked crystal structure and exhibits semiconducting transport behavior instead\cite{xiaoHeavyFermionBehaviorTunable2026}. Therefore, the \a3913\ family provides a promising platform for exploring the interplay between geometric frustration, electronic correlation, and dimensionality. However, despite the rapidly growing experimental interest, the microscopic electronic structure, magnetic properties and the role of geometric frustration in these materials remain theoretically unexplored.

In this article, we combine density functional theory (DFT) calculations and Monte Carlo simulations to investigate the electronic structure and magnetic properties of the \a3913\ family. A remarkably robust vortex-like magnetic ground state arising from the frustrated geometry is discovered for \cs3913. Owing to the quasi-two-dimensional nature of the system, the vortex phase hosts a manifold of nearly degenerate magnetic configurations together with Dirac-cone dispersions, van Hove singularities, and flat-band electronic states. The complicated magnetic order can be captured by a simple Heisenberg model based on local moments on V-sites, which reproduces the experimentally observed weak dependence of the transition temperature on external field. Replacing Cs with Rb leads to broken $C_3$ symmetry due to AB-stacking instead of AA-stacking, resulting in a small gap at the Fermi level in \rb3913.

\section{Results}

\subsection{Magnetic Ground State}

We first explore the magnetic ground state using DFT calculations\footnote{
All calculations are done assuming full spinor wavefunctions without spin-orbit coupling unless otherwise specified. Such set-up allows noncollinear magnetic configurations without considering spin-orbit coupling. For other detailed calculation parameters, please refer to Supplementary Materials.}. Since the experiments suggest weak inter-layer coupling as well as large in-plane lattice constants, the search for the ground state configuration is constrained in a single unit-cell. We consider only the magnetic orders that preserve the $C_3$ symmetries and mirror symmetry. In collinear cases, it requires ferromagnetic (FM) alignment of the moments within each triangle; while in the coplanar cases, it requires the moments within each triangle to form either vortex or anti-vortex (FIG. \ref{fig:fig1}b). In the vortex/anti-vortex case, the moments on neighboring sites are rotated 120$^{\circ}$ clockwise/counter-clockwise, when the triangle is traversed clockwise. In addition, the magnetic moments on nearest neighboring Cr-sites between two large triangles, as well as between the small and large triangles, are predominantly anti-parallel aligned. To evaluate the inter-layer coupling strengths, we also include the inter-layer antiferromagnetic (AFM) configuration, which was proposed previously as the low-energy state\cite{zhaoCs3913_NewVanadiumBased2026}. All considered initial configurations are shown in FIG. \ref{fig:fig1}. 

\begin{table}[htp]
\caption{
Optimized lattice constants, magnetic moments, and relative energies of different magnetic configurations for \cs3913. The NCL-L/mixed-VA configurations are unstable and automatically converge to the vortex/anti-vortex phases, respectively.\label{tab1}}
\begin{ruledtabular}
\begin{tabular}{c|ccccc}
Config. &
$a$ (\AA) &
$c$ (\AA) &
$m_L$ ($\mu_B$) &
$m_S$ ($\mu_B$) &
$\Delta E$ (meV/f.u.) \\
\hline
Vortex  & 10.180 & 8.581 & 1.68 & 1.03 & 0 \\
Anti-Vortex & 10.180 & 8.581 & 1.68 & 1.03 & 0 \\
FM-LS             & 10.198 & 8.508 & 1.53 & 0.06 & 286.75 \\
A-type            & 10.199 & 8.504 & 1.53 & 0.03 & 287.41 \\
AFM-LL            & 10.246 & 8.429 & 1.78 & 0    & 360.73 \\
NCL-S             & 10.049 & 8.656 & 0.11 & 0.97 & 1201.46 \\
NCL-L             & \multicolumn{5}{c}{converges to vortex} \\
all Mixed-VA      & \multicolumn{5}{c}{converges to anti-vortex} \\
FM-S              & 10.036 & 8.671 & 0.05 & 0.64 & 1264.92 \\
NM                & 10.009 & 8.675 & --   & --   & 1332.20 \\
\end{tabular}
\end{ruledtabular}
\end{table}

The results of DFT calculations for all these configurations are summarized in TABLE \ref{tab1}. The lattice constants and internal coordinates are fully relaxed in all cases. The inter-layer coupling is negligible, as the A-type AFM is almost degenerate with FM configuration. They are $\sim$300 meV/f.u. higher in energy than the vortex phase, which is the ground state of \cs3913. It is worth noting that both vortex and anti-vortex configurations belong to the same spin space group (SSG) in the nonrelativistic limit\cite{10.1098/rspa.1966.0211, litvinSpinGroups1974, litvinSpinPointGroups1977, xiaoSpinSpaceGroups2024, chenEnumerationRepresentationTheory2024, jiangEnumerationSpinSpaceGroups2024}, and they have identical total energies, lattice parameters, and magnitudes of local moments in calculations without spin-orbit coupling (SOC), consistent with the Heisenberg-model analysis discussed below. The SSG ensures a vanishing net magnetization in both configurations. Once SOC is included, however, the spin and lattice degrees of freedom become coupled and the two configurations become distinguishable under magnetic space-group (MSG). The vortex state belongs to ($P\bar{6}'2m'$) (No.~189.224), whereas the anti-vortex state belongs to the lower-symmetry ($Am'm'2$) (No.~38.191) magnetic space group. Although a vanishing net magnetization is still enforced by MSG of the vortex state, it is not the case for the anti-vortex state. Therefore, the anti-vortex configuration belongs to the AFM(SOM) class proposed by Liu \textit{et al.}\cite{liuSymmetryClassificationMagnetic2026a}, namely an antiferromagnetic state in the nonrelativistic limit in which a finite net spin magnetization is symmetry-allowed to emerge only through SOC. Correspondingly, SOC lifts the vortex--anti-vortex degeneracy, raising the total energy of the anti-vortex state by approximately 53 meV/f.u. relative to the vortex state. The vortex phase remains robust upon inclusion of SOC, and the in-plane magnetic anisotropy energy turns out to be small, as suggested by the negligible variation ($\sim$ 1.5 meV/f.u.) of the total energies (obtained with SOC) when all moments rotate the same angle around $c$-axis. The fully relaxed lattice constant $a=10.180$ \AA\ is in good agreement with experimental value ($a_{\mathrm{exp}}=10.137$ \AA), whereas $c=8.580$\AA\ is slightly larger than $c_{\mathrm{exp}}=8.147$ \AA. Such behavior is frequently observed in quasi-2D materials such as transition-metal dichalcogenide or graphite\cite{rydbergVanWaalsDensity2003, joshiShortrangeChargeDensity2019a, spanuNatureStrengthInterlayer2009}, suggesting weak inter-layer coupling and is consistent with experimental observation\cite{xiaoHeavyFermionBehaviorTunable2026, zhaoCs3913_NewVanadiumBased2026}. Under pressure, the variation of the lattice constants agrees well with the experimental trend [Fig.~\ref{fig:fig2}(a)]\cite{zhaoCs3913_NewVanadiumBased2026}. In particular, the $c$ axis contracts more rapidly than the in-plane lattice constant $a$ with increasing pressure, further reflecting the pronounced structural anisotropy and quasi-two-dimensional character of Cs$_3$V$_9$Te$_{13}$.

\subsection{Heisenberg Model Analysis}
To understand the vortex phase, we employ a Heisenberg model with nearest-neighbor-only interactions:
\begin{eqnarray*}
 H_{\mathrm{NN}} & = & \frac{J_S}{2} \sum_{i, j \in \bigtriangleup_{\mathrm{S}}} \mathbf{S}_i \cdot \mathbf{S}_j + \frac{J_L}{2} \sum_{i, j \in \bigtriangleup_{\mathrm{L}}} \mathbf{S}_i \cdot \mathbf{S}_j \\
                 & + & \frac{J_{LL}}{2} \sum_{\langle i, j \rangle} \mathbf{S}_i \cdot \mathbf{S}_j + \frac{J_{SL}}{2} \sum_{[i, j]} \mathbf{S}_i \cdot \mathbf{S}_j
\end{eqnarray*}

where $i, j$ in the first 2 terms take sites within the same small (large) triangle denoted by $\bigtriangleup_{\mathrm{S}}$ ($\bigtriangleup_{\mathrm{L}}$), and $\langle i,j \rangle$ ($[i,j]$) denote nearest-neighboring sites between 2 large triangles (between small and large triangles). $J_S$, $J_L$, $J_{LL}$ and $J_{SL}$ are the corresponding exchange interactions, respectively (FIG. \ref{fig:fig1}a).

Due to the $C_3$ symmetry for each triangle (large or small) around the axis through its center, the classical total energies contributed by the vortex and anti-vortex configurations from each triangle are both $-\frac{3}{2}JS^2$, and the FM configuration contributes $3JS^2$, where $(J, S)$ can be $(J_L, S_L)$ or $(J_S, S_S)$ for large or small triangles, respectively. To facilitate the following discussion, we define the angles formed by the $[110]$ direction and the magnetic moment of V$^1$, V$^4$, and V$^7$ atoms [as shown in FIG. \ref{fig:fig1}(a)] as $\theta_1$, $\theta_2$ and $\theta_3$, respectively. 

It can be shown that if all exchange interactions are AFM ($J>0$), all triangles must be either vortex or anti-vortex for the ground state configuration, and $\theta_2-\theta_1-\pi=\theta_1-\theta_3+\pi=\Delta\theta$ (please refer to SI for details). The second condition simply means that the angles formed by moments on nearest neighboring small/large triangles must be the same, and is also consistent with the mirror symmetry relating the 2 large triangles (FIG. \ref{fig:fig1}a). These conditions put very strong constraints on the ground state magnetic configuration of these compounds, such that the spin configuration of these triangles are ``interlocked", i.e.: the directions of all moments are indeed coupled such that fixing the spin configuration of 1 triangle propagates and fixes the rest. In this case, the classical total energy for each unit cell $E_{\mathrm{NN}}$ is reduced to:

\begin{equation}
\begin{aligned}
E_{\mathrm{NN}} = - \Bigg[& \frac{3}{2} J_S |S_1|^2 + 3 J_L |S_2|^2 + 
\\ & 3 J_{SL}|S_1||S_2| f(\Delta\theta, \alpha)
\Bigg].
\end{aligned}
\label{eq:HNN2}
\end{equation}
with $\alpha = \frac{J_{LL}|S_2|}{J_{SL}|S_1|}$ and
\begin{equation}
f(\Delta\theta, \alpha)
= 2\cos\Delta\theta +  \alpha \cos\left(2\Delta\theta\mp\frac{\pi}{3}\right)
\end{equation}
, where the vortex phase takes the minus sign and the anti-vortex phase takes the plus sign. Thus, the ground state energies of vortex and anti-vortex phases must be identical in this simple model, and their canting angles are related by $\Delta\theta_{\mathrm{vortex}}=-\Delta\theta_{\mathrm{anti-vortex}}$.

\begin{figure*}[t]
\includegraphics[width=\textwidth]{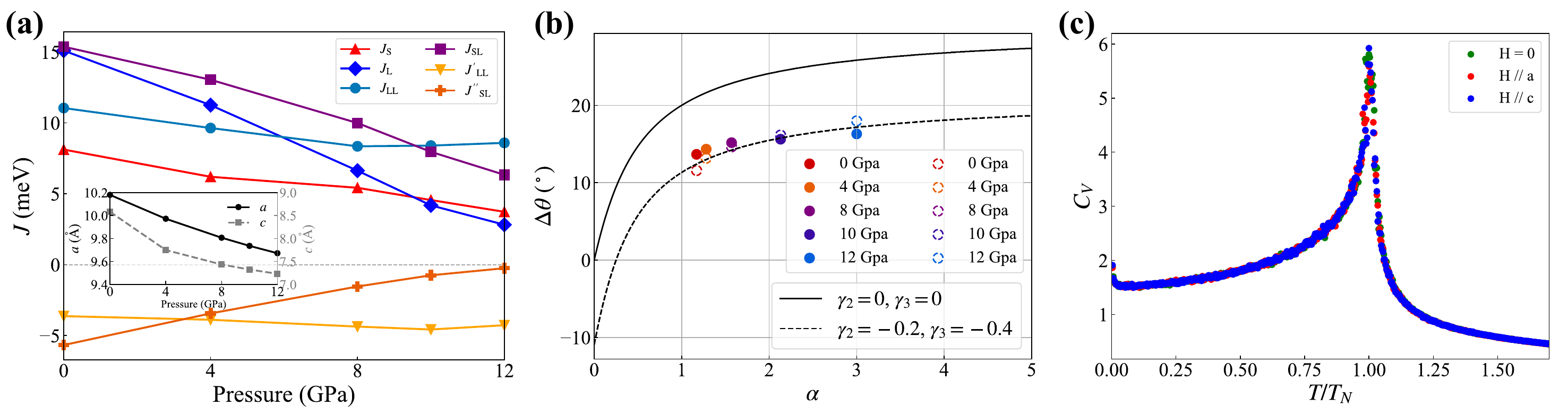}
\caption{(a) Pressure dependence of the lattice constants and the dominant magnetic exchange interactions in Cs$_3$V$_9$Te$_{13}$. (b) Relative canting angle $\Delta\theta$ obtained from DFT structural relaxation at different pressures, together with the classical Heisenberg-model prediction as a function of $\alpha=J_{LL}|S_2|/(J_{SL}|S_1|)$. Solid symbols denote the values obtained directly from DFT structural relaxation at different pressures, whereas open symbols represent the canting angles predicted from the exchange interactions extracted at the corresponding pressures. The solid curve includes only the nearest-neighbor exchange interactions, whereas the dashed curve includes both nearest- and next-nearest-neighbor interactions with $f_{\rm NNN}(\Delta\theta,\alpha)=-2\gamma_2\cos\left(\Delta\theta+\frac{\pi}{3}\right)-\alpha\gamma_3\cos(2\Delta\theta)$. (c) Temperature dependence of the specific heat obtained from parameter-free Monte Carlo simulations under magnetic fields applied along the $a$ and $c$ axes, together with the zero-field result.}
\label{fig:fig2}
\end{figure*}

The relation between the optimal $\Delta\theta$ minimizing $E_{\mathrm{NN}}$ and $\alpha$ for $H_{\mathrm{NN}}$ is shown in Fig.~\ref{fig:fig2}(b). If the in-plane anisotropy is small, such relation guarantees the emergence of a continuous rotational degree of freedom associated with the magnetic moment of the small triangle. Consequently, the vortex order is not an isolated magnetic configuration but belongs to a manifold of nearly degenerate states generated by geometric frustration. Interestingly, the above conclusions remain valid even if long-range (e.g. next-nearest-neighboring) interactions are included in the Heisenberg model (see Supplemental Materials for detail). 

\begin{figure*}[t]
\includegraphics[width=\textwidth]{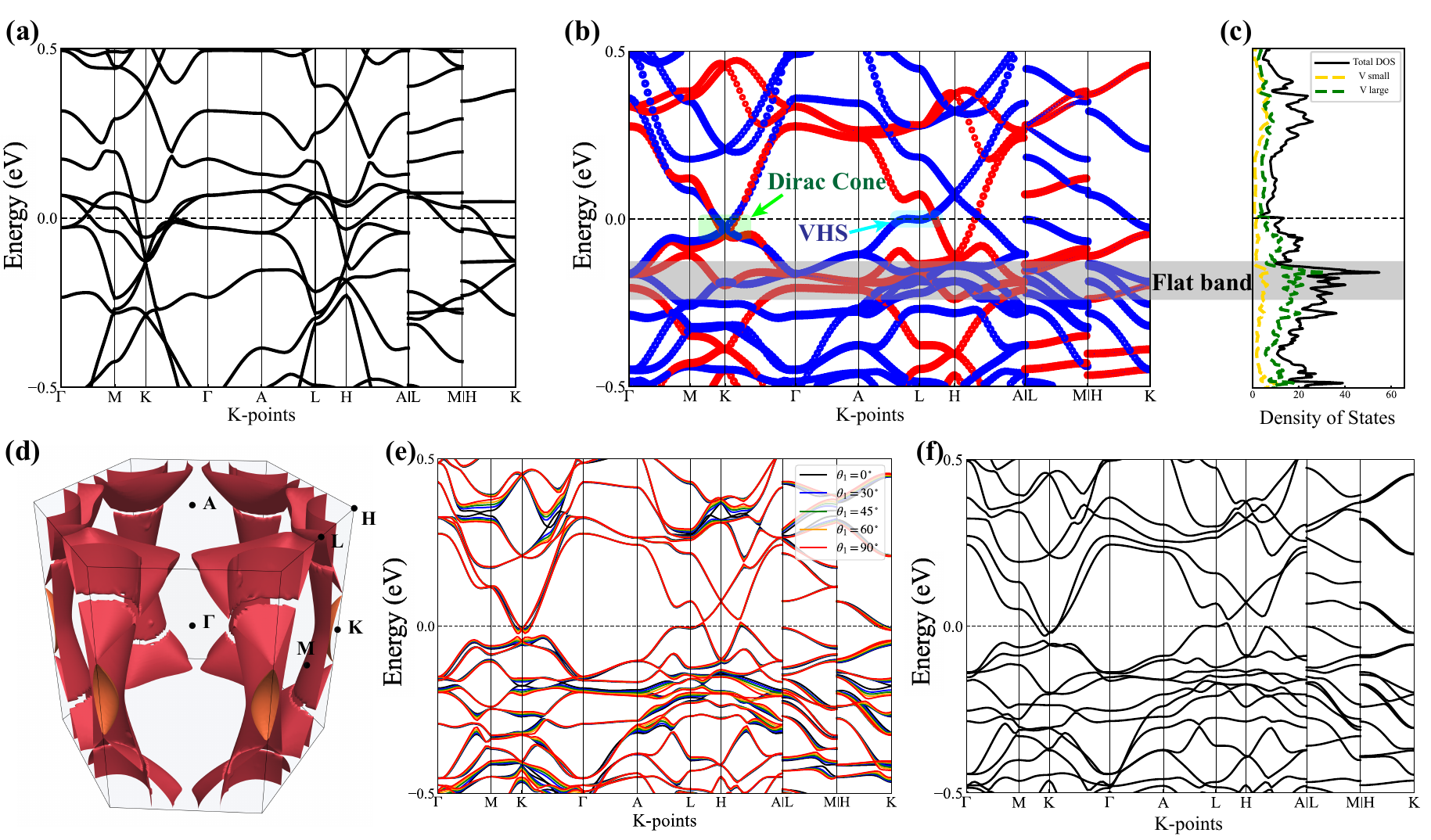}
\caption{
(a) Electronic band structure of the nonmagnetic (NM) state of Cs$_3$V$_9$Te$_{13}$. (b),(c) Orbital-projected band structure and density of states of the vortex phase without SOC. Red (blue) circles denote the weights of the V $d_{yz}+d_{xz}$ ($d_{x^2-y^2}+d_{xy}+d_{z^2}$) orbitals. In the DOS, green and gold dashed curves correspond to V atoms belonging to the large and small triangles, respectively. (d) Brillouin zone and Fermi surface of the vortex phase. (e) SOC band structures of the vortex states with different $\theta_1$ angles. (f) Electronic band structure of the anti-vortex configuration with SOC.
}
\label{fig:fig3}
\end{figure*}

The magnetic exchange interactions at ambient pressure were extracted using TB2J\cite{PhysRevB.91.224405, ANTROPOV1997336, HE2021107938} from the Wannier-projected Hamiltonian\cite{wannier,pizziWannier90CommunityCode2020} of the vortex configuration. The inter-layer exchange interactions are extremely weak (all smaller than 0.1~meV), which is consistent with the DFT total energies, and indicates a pronounced quasi-two-dimensional magnetic character in the A$_3$V$_9$Te$_{13}$ system. The intra-triangle exchange interactions within the large and small triangles are $J_L=15.10$~meV and $J_S=8.12$~meV, respectively. The nearest-neighbor interactions between the small and large triangles and between adjacent large triangles are $J_{SL}=15.37$~meV and $J_{LL}=11.05$~meV, respectively. These value yield $\alpha\approx1.2$ at ambient pressure. In this case, the simple nearest neighbor model $H_{\mathrm{NN}}$ suggests the optimal canting angle $\Delta\theta\approx20^{\circ}$, while our first-principles calculations yield $\Delta\theta=13.7^{\circ}$. 

To demonstrate the validity of the spin model, we have also calculated the exchange interactions between 0 GPa and 12 GPa (FIG. \ref{fig:fig2}a). With increasing pressure, $J_{SL}$ decreases substantially, while $J_{LL}$ is reduced much more slowly. Consequently, the dimensionless interaction ratio $\alpha$ monotonically increases to approximately 3 at 12 GPa. The optimal canting angles obtained from nearest neighboring Heisenberg model exhibit a trend similar to the first-principles values (FIG. \ref{fig:fig2}b). 

For a quantitative comparison, we then include next-nearest neighbor corrections. The next-nearest-neighbor interactions $J'_{LL}$ and $J''_{SL}$ provide non-negligible corrections, while other longer-range interactions are relatively weak, with magnitudes below 1~meV(The geometrical definitions of these longer-range exchange paths are shown in Fig.~S1). Neglecting $J'_{SL}$ and $J''_{LL}$, the total energy correction up to the next-nearest neighbor is:
\begin{equation}
E'_{\mathrm{NNN}}=-3|S_1||S_2|J_{SL}f_{\rm NNN}(\Delta\theta,\alpha,\gamma_2,\gamma_3)
\end{equation}
with $\gamma_2=\frac{J''_{SL}}{J_{SL}}$, $\gamma_3=\frac{J'_{LL}}{J_{LL}}$, and
$$f_{\rm NNN}(\Delta\theta,\alpha) = -2\gamma_2\cos\left(\Delta\theta\pm\frac{\pi}{3}\right)-\alpha\gamma_3\cos(2\Delta\theta)$$
Since the fitted $\gamma_2$ and $\gamma_3$ vary only weakly with pressure, we approximate them by the average values $\gamma_2=-0.2$ and $\gamma_3=-0.4$. In Fig.~\ref{fig:fig2}(b), the open symbols represent the optimal canting angles predicted using the pressure-dependent fitted values of $\gamma_2$ and $\gamma_3$, which agree closely with the DFT-relaxed values shown by the solid symbols. These open symbols also fall near the dashed curve obtained with the average $\gamma_2$ and $\gamma_3$, confirming that the pressure dependence of the next-nearest-neighbor corrections is weak.

\subsection{Monte Carlo Simulations}

Using the exchange parameters obtained from TB2J, including all isotropic and anisotropic magnetic interactions between V atoms up to 7.7~\AA, we performed classical Monte Carlo simulations\cite{evansAtomisticSpinModel2014a}. The calculated magnetic ground state is consistent with the vortex-like configuration predicted by the Heisenberg-model analysis and the DFT calculations (Fig.~S2a). In addition, the magnetic transition temperature remains nearly unchanged regardless of the direction or magnitude of the applied magnetic field (FIG. \ref{fig:fig2}c). We note that such behavior is consistent with experimental observation\cite{xiaoHeavyFermionBehaviorTunable2026,zhaoCs3913_NewVanadiumBased2026,liuCs3913CorrelatedElectron2026}, and is most likely to be an intrinsic consequence of the vortex magnetic order. Furthermore, both our simulations and the experiments consistently observe a more pronounced variation of the susceptibility for $H\parallel ab$ than for $H\parallel c$ upon approaching $T_N$. Similar susceptibility behavior has also been reported in the vortex-magnetic compound CrRhAs\cite{shangNoncollinearAntiferromagneticStructure2026a}, further supporting the vortex nature of the magnetic order in \cs3913.

\subsection{Electronic Structures}

We next investigate the electronic structure of Cs$_3$V$_9$Te$_{13}$. The electronic band structure of the nonmagnetic state is shown in Fig.~\ref{fig:fig3}(a), while the band structure, density of states, and Fermi surface of the vortex-like phase without SOC are presented in Fig.~\ref{fig:fig3}(b)--(d). The calculated Fermi surface also reveals pronounced quasi-two-dimensional electronic characteristics in \cs3913. Near the Fermi level, Dirac-like dispersions are found around the K point. The breaking of time-reversal symmetry lifts the fourfold Dirac degeneracy and splits the original Dirac point into two spin-polarized band crossings, which are predominantly derived from the V $d_{yz}+d_{xz}$ and $d_{x^2-y^2}+d_{xy}+d_{z^2}$ orbitals, respectively. Similar split band-crossings are also present around the H point. The larger splitting around H shifts the van Hove singularity associated with the $d_{x^2-y^2}+d_{xy}+d_{z^2}$ orbitals to the vicinity of the Fermi level near the L point. In addition, the split bands produce crossing points near the Fermi level along the L-H and H-A paths. A nearly flat band located about 200~meV below the Fermi level is mainly contributed by V atoms belonging to the large triangles. The coexistence of Dirac-like dispersions, van Hove singularities, and flat bands reflects the vestige from parent kagome lattice.

Once SOC is included, it generally gaps out the Weyl-like crossings around K and H [Fig.~\ref{fig:fig3}(e)]. Nevertheless, our calculations identify 6 pairs of Weyl points in the BZ for the vortex phase. Therefore, it remains metallic in the presence of SOC. We have performed calculations with different $\theta_1$ angles, ranging from 0$^{\circ}$ to 90$^{\circ}$. The small in-plane anisotropy is also reflected in the electronic structure, as they all produce very similar band dispersions. A similar conclusion can also be drawn for the electronic structure of the anti-vortex phase. Without SOC, the band structure of the anti-vortex phase is identical to that of the vortex phase. With SOC, its low-energy band structure close to the Fermi level as well as its orbital characters remain qualitatively similar to the vortex phase as well [FIG. \ref{fig:fig3}(f)]. Due to the reduced symmetry in the anti-vortex phase, 4 instead of 6 pairs of Weyl points can be identified, although the system is still metallic.

\begin{figure*}[t]
\includegraphics[width=\textwidth]{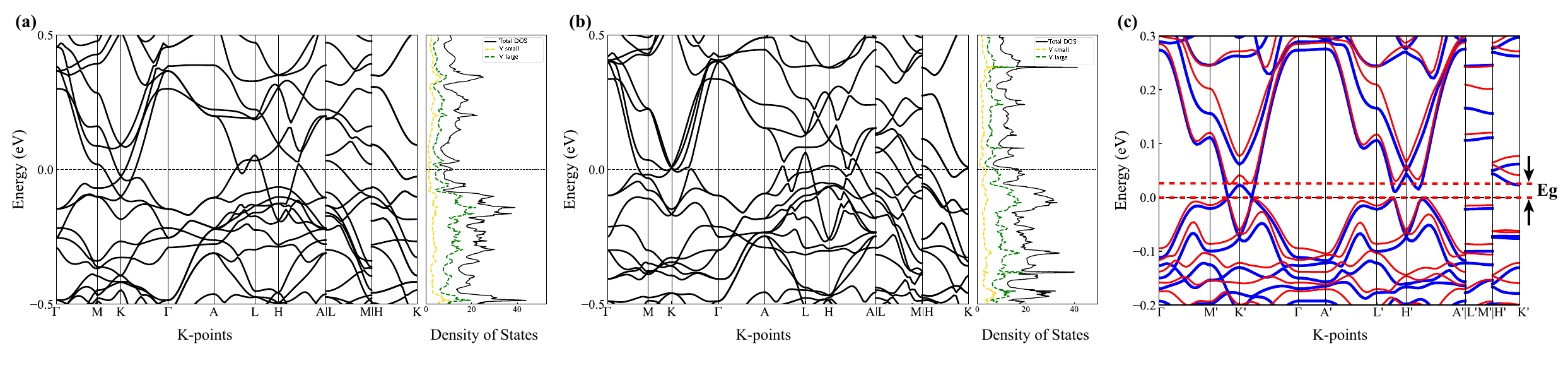}
\caption{
(a)-(b) Electronic band structures and densities of states of the vortex phase in Cs$_3$V$_9$Te$_{13}$ at 4 and 8~GPa, respectively, calculated without SOC. (c) Electronic band structures of the vortex phase in Rb$_3$V$_9$Te$_{13}$. Blue lines and red lines denote the band structures without and with SOC, respectively.
}
\label{fig:fig4}
\end{figure*}

We further calculated the pressure evolution of the electronic structure of Cs$_3$V$_9$Te$_{13}$. The band structures and densities of states at 4~GPa and 8~GPa are shown in Fig.~\ref{fig:fig4}(a) and Fig.~\ref{fig:fig4}(b), respectively. With increasing pressure, the flat band located below the Fermi level gradually shifts toward the Fermi energy, leading to an enhanced density of states in its vicinity. Such enhancement in the low-energy electronic states is in agreement with experimental observation of enhancement of metallicity\cite{zhaoCs3913_NewVanadiumBased2026}, and suggests possible variation in electronic correlation\cite{liuCs3913CorrelatedElectron2026}.

In contrast to \cs3913, replacing Cs with Rb leads to substantial stacking difference, and modification to electronic structure thereafter. \rb3913\ adopts an AB-stacking order instead of AA-stacking, and the crystal symmetry is reduced from the hexagonal space group $P\bar{6}2m$ (No.~189) to the orthorhombic space group $Cmcm$ (No.~63)\cite{xiaoHeavyFermionBehaviorTunable2026}. Consequently, the $C_3$ rotational symmetry present in Cs$_3$V$_9$Te$_{13}$ is removed. At ambient pressure, the inter-layer distance is $\sim$8 \AA, similar to \cs3913. Therefore, we assume that the inter-layer magnetic exchange is also weak, and the ground state is also vortex-like configuration (FIG. \ref{fig:fig3}a). Since the $C_3$ symmetry is lifted by the stacking order, the band crossings and Weyl points are no longer present and completely removed in \rb3913\ (FIG. \ref{fig:fig4}(c)). In addition, SOC opens a direct gap of approximately 30 meV, resulting in a semiconducting electronic structure at ambient pressure, consistent with experimental observations\cite{xiaoHeavyFermionBehaviorTunable2026}. 

\section{Discussion}

For the A$_3$V$_9$Te$_{13}$ family, the fitted interlayer exchange interaction $J_c$ indicates that the magnetic coupling between neighboring V layers is extremely weak. Such weak interlayer coupling is also reflected in the electronic structure, where both electronic structure calculations and experimental measurements reveal pronounced quasi-two-dimensional characteristics of the Fermi surface. Furthermore, our classical Heisenberg-model analysis and DFT calculations demonstrate that the in-plane magnetic anisotropy energies are very small as well. Therefore, it is possible for each layer to form a 2D magnetic order without an ordered arrangement of moment directions across different layers at higher temperatures. A long-range order in $c$-direction may take place at a much lower temperature $\sim J_c$. In addition, we note that classical Monte Carlo simulations yield a magnetic transition temperature of $T_N \approx 292$~K, significantly higher than the experimentally observed transition temperature of $47$~K. This discrepancy may arise from two factors. First, the exchange parameters extracted from DFT calculations are likely overestimated, a well-known effect in studies of 3d-materials\cite{PhysRevB.78.224517,Rocquefelte:2012aa,daglum2025abinitiocalculationmagneticexchange}. Second, the classical Monte Carlo approach neglects quantum fluctuations that can substantially suppress long-range antiferromagnetic order in frustrated magnetic systems. Such a reduction of the ordering temperature is expected to be particularly pronounced in quasi-two-dimensional frustrated magnets, where quantum fluctuations can strongly suppress long-range magnetic order. Despite these discrepancies, the vortex-like magnetic phase appears to be robust and its transition temperature is nearly independent of external magnetic field, consistent with experimental observations.

In conclusion, we have systematically investigated the electronic structures and magnetic properties of the A$_3$V$_9$Te$_{13}$ family by combining classical Heisenberg-model analysis with first-principles calculations. Our results reveal the existence of a remarkably robust ``interlocked" vortex/anti-vortex magnetic order. The vortex and anti-vortex states are exactly degenerate in the nonrelativistic limit, whereas SOC lifts the chirality degeneracy and selects the vortex state. The vortex ground state is characterized by nearly anti-parallel alignment between nearest neighboring inter-triangle Cr-sites within the same pentagon-rings.  Owing to the small in-plane anisotropy and quasi-two-dimensional nature of the A$_3$V$_9$Te$_{13}$ lattice, the system hosts a manifold of nearly degenerate vortex/anti-vortex magnetic states together with Dirac-cone dispersions, van Hove singularities, and flat-band electronic structures. Pressure simultaneously tunes the exchange-interaction ratio, the magnetic canting angle, and the low-energy electronic density of states. Monte Carlo simulations reproduce the experimentally observed weak magnetic-field dependence of the magnetic transition temperature, highlighting the unusual nature of the vortex phase. For Rb$_3$V$_9$Te$_{13}$, the AB stacking removes the threefold rotational symmetry present in the Cs compound, and SOC opens a small semiconducting gap under ambient pressure. Our results establish A$_3$V$_9$Te$_{13}$ as a promising platform for exploring frustrated magnetism and pressure-tunable electronic states in vortex magnetic systems.

\section{Methods}
\subsection{First-principles calculations}
All first-principles calculations were performed using the projector augmented-wave (PAW) method as implemented in the VASP package\cite{method:pawvasp}. The exchange-correlation energy was treated within the generalized gradient approximation (GGA) using the Perdew-Burke-Ernzerhof (PBE) functional\cite{method:pbe}. A plane-wave cutoff energy of 500~eV was employed, and the Brillouin zone was sampled using an $8\times8\times8$ $k$-point mesh. All structures under different magnetic configurations were fully relaxed within the noncollinear framework using identical computational settings.

\subsection{Monte Carlo simulations}
Classical Monte Carlo simulations were performed using the Vampire package\cite{evansAtomisticSpinModel2014a} based on the magnetic exchange parameters within distance $|R_{ij}|<7.7$~\AA, including both isotropic and anisotropic interactions. A $20\times20\times1$ supercell containing 3600 magnetic atoms was adopted. $1.0\times10^5$ measurement steps were taken after the equilibrium was achieved. For field-dependent simulations, magnetic fields of 100~T were applied along the $a$ and $c$ directions.

\section{Data availability}
The data that support the findings of this study are available from the corresponding author upon reasonable request.

\section{Code availability}
No custom code was used in this study.

\section{Acknowledgments}
This work has been supported by the National Natural Science Foundation of China (Nos. 12350710785, 12674174, \& 12574163) and the National Key R\&D Program of China (Nos. 2024YFA1408303, 2022YFA1402202, \& 2022YFA402701). The calculations were performed on the Quantum Many-Body Computing Cluster at Zhejiang University and High Performance Computing Center of Hangzhou Normal University.

\section{Author contributions}
C.C. and X.W. initialized the project. Z.L. carried out all the calculations and data analysis under supervision of C.C.. H.-Q. Y. and X.W. participated the discussion, and contributed to the interpretation of the numerical results. Z.L. and C.C. drafted the manuscript with input from all authors. 

\section{Competing interests}
The authors declare no competing interests.

\bibliography{ref}

@article{PhysRevB.78.224517,
	author = {Ma, Fengjie and Lu, Zhong-Yi and Xiang, Tao},
	doi = {10.1103/PhysRevB.78.224517},
	issue = {22},
	journal = {Phys. Rev. B},
	month = {Dec},
	numpages = {6},
	pages = {224517},
	publisher = {American Physical Society},
	title = {Arsenic-bridged antiferromagnetic superexchange interactions in LaFeAsO},
	url = {https://link.aps.org/doi/10.1103/PhysRevB.78.224517},
	volume = {78},
	year = {2008}}

@misc{daglum2025abinitiocalculationmagneticexchange,
	archiveprefix = {arXiv},
	author = {Umit Dogan Daglum and Maria Stamenova and Ersoy Sasioglu and Stefano Sanvito},
	eprint = {2510.11603},
	primaryclass = {cond-mat.mtrl-sci},
	title = {Ab-initio calculation of magnetic exchange interactions using the spin-spiral method in VASP: Self-consistent versus magnetic force theorem approaches},
	url = {https://arxiv.org/abs/2510.11603},
	year = {2025}}

@article{Rocquefelte:2012aa,
	author = {Rocquefelte, Xavier and Schwarz, Karlheinz and Blaha, Peter},
	date = {2012/10/22},
	doi = {10.1038/srep00759},
	id = {Rocquefelte2012},
	isbn = {2045-2322},
	journal = {Scientific Reports},
	number = {1},
	pages = {759},
	title = {Theoretical Investigation of the Magnetic Exchange Interactions in Copper(II) Oxides under Chemical and Physical Pressures},
	url = {https://doi.org/10.1038/srep00759},
	volume = {2},
	year = {2012}}

@article{wannier,
	author = {Arash A. Mostofi and Jonathan R. Yates and Young-Su Lee and Ivo Souza and David Vanderbilt and Nicola Marzari},
	doi = {https://doi.org/10.1016/j.cpc.2007.11.016},
	issn = {0010-4655},
	journal = {Computer Physics Communications},
	number = {9},
	pages = {685-699},
	title = {wannier90: A tool for obtaining maximally-localised Wannier functions},
	url = {https://www.sciencedirect.com/science/article/pii/S0010465507004936},
	volume = {178},
	year = {2008}}

@article{HE2021107938,
	author = {Xu He and Nicole Helbig and Matthieu J. Verstraete and Eric Bousquet},
	doi = {https://doi.org/10.1016/j.cpc.2021.107938},
	issn = {0010-4655},
	journal = {Computer Physics Communications},
	pages = {107938},
	title = {TB2J: A python package for computing magnetic interaction parameters},
	url = {https://www.sciencedirect.com/science/article/pii/S0010465521000679},
	volume = {264},
	year = {2021}}

@article{ANTROPOV1997336,
	author = {V.P. Antropov and M.I. Katsnelson and A.I. Liechtenstein},
	doi = {https://doi.org/10.1016/S0921-4526(97)00203-2},
	issn = {0921-4526},
	journal = {Physica B: Condensed Matter},
	note = {Proceedings of the Yamada Conference XLV, the International Conference on the Physics of Transition Metals},
	pages = {336-340},
	title = {Exchange interactions in magnets},
	url = {https://www.sciencedirect.com/science/article/pii/S0921452697002032},
	volume = {237-238},
	year = {1997}}

@article{PhysRevB.91.224405,
	author = {Korotin, Dm. M. and Mazurenko, V. V. and Anisimov, V. I. and Streltsov, S. V.},
	doi = {10.1103/PhysRevB.91.224405},
	issue = {22},
	journal = {Phys. Rev. B},
	month = {Jun},
	numpages = {7},
	pages = {224405},
	publisher = {American Physical Society},
	title = {Calculation of exchange constants of the Heisenberg model in plane-wave-based methods using the Green's function approach},
	url = {https://link.aps.org/doi/10.1103/PhysRevB.91.224405},
	volume = {91},
	year = {2015}}

@article{spanuNatureStrengthInterlayer2009,
	author = {Spanu, Leonardo and Sorella, Sandro and Galli, Giulia},
	doi = {10.1103/PhysRevLett.103.196401},
	issue = {19},
	journal = {Phys. Rev. Lett.},
	month = {Nov},
	numpages = {4},
	pages = {196401},
	publisher = {American Physical Society},
	title = {Nature and Strength of Interlayer Binding in Graphite},
	url = {https://link.aps.org/doi/10.1103/PhysRevLett.103.196401},
	volume = {103},
	year = {2009}}

@article{joshiShortrangeChargeDensity2019a,
	author = {Joshi, Jaydeep and Hill, Heather M. and Chowdhury, Sugata and Malliakas, Christos D. and Tavazza, Francesca and Chatterjee, Utpal and Hight Walker, Angela R. and Vora, Patrick M.},
	doi = {10.1103/PhysRevB.99.245144},
	issue = {24},
	journal = {Phys. Rev. B},
	month = {Jun},
	numpages = {9},
	pages = {245144},
	publisher = {American Physical Society},
	title = {Short-range charge density wave order in $2H\text{\ensuremath{-}}\mathrm{T}\mathrm{a}{\mathrm{S}}_{2}$},
	url = {https://link.aps.org/doi/10.1103/PhysRevB.99.245144},
	volume = {99},
	year = {2019}}

@article{rydbergVanWaalsDensity2003,
	author = {Rydberg, H. and Dion, M. and Jacobson, N. and Schr\"oder, E. and Hyldgaard, P. and Simak, S. I. and Langreth, D. C. and Lundqvist, B. I.},
	doi = {10.1103/PhysRevLett.91.126402},
	issue = {12},
	journal = {Phys. Rev. Lett.},
	month = {Sep},
	numpages = {4},
	pages = {126402},
	publisher = {American Physical Society},
	title = {Van der Waals Density Functional for Layered Structures},
	url = {https://link.aps.org/doi/10.1103/PhysRevLett.91.126402},
	volume = {91},
	year = {2003}}

@article{zhaoCs3913_NewVanadiumBased2026,
	author = {Zhao, Zhen and Sun, Jianping and Yi, Xin-Wei and Liu, Tong and Zhu, Lin and Ma, Hanming and Wang, Ruwen and Zhang, Hua and Zhang, Yuhang and Liu, Haisen and Ma, Xiaoli and Guo, Hui and Lin, Xiao and Bao, Lihong and Yu, Xiaohui and Zhou, Wu and Cheng, Jinguang and Su, Gang and Yang, Haitao and Gao, Hong-Jun},
	doi = {10.1088/0256-307X/43/6/060703},
	journal = {Chinese Physics Letters},
	month = {jun},
	number = {6},
	pages = {060703},
	publisher = {Chinese Physical Society and IOP Publishing Ltd},
	title = {Cs$_3$V$_9$Te$_13$: A New Vanadium-Based Material with a Reuleaux-Triangle-Like Lattice and a Possible Phase Transition near 48 K},
	url = {https://doi.org/10.1088/0256-307X/43/6/060703},
	volume = {43},
	year = {2026}}

@article{yuChiralSuperconductingPhase2012,
	author = {Yu, Shun-Li and Li, Jian-Xin},
	journal = {PHYSICAL REVIEW B},
	title = {Chiral Superconducting Phase and Chiral Spin-Density-Wave Phase in a {{Hubbard}} Model on the Kagome Lattice},
	year = {2012}}

@article{kagome_1,
	author = {Sy{\^o}zi, Itiro},
	doi = {10.1143/ptp/6.3.306},
	eprint = {https://academic.oup.com/ptp/article-pdf/6/3/306/5239621/6-3-306.pdf},
	issn = {0033-068X},
	journal = {Progress of Theoretical Physics},
	month = jun,
	number = {3},
	pages = {306--308},
	title = {Statistics of Kagom\'e Lattice},
	volume = {6},
	year = 1951}

@article{mielkeFerromagneticGroundStates1991,
	author = {Mielke, A.},
	doi = {10.1088/0305-4470/24/2/005},
	issn = {0305-4470},
	journal = {Journal of Physics A: Mathematical and General},
	langid = {english},
	month = jan,
	number = {2},
	pages = {L73},
	title = {Ferromagnetic Ground States for the {{Hubbard}} Model on Line Graphs},
	urldate = {2026-06-10},
	volume = {24},
	year = 1991}

@article{yinTopologicalKagomeMagnets2022,
	author = {Yin, Jia-Xin and Lian, Biao and Hasan, M. Zahid},
	copyright = {2022 Springer Nature Limited},
	doi = {10.1038/s41586-022-05516-0},
	issn = {1476-4687},
	journal = {Nature},
	langid = {english},
	month = dec,
	number = {7941},
	pages = {647--657},
	publisher = {Nature Publishing Group},
	title = {Topological Kagome Magnets and Superconductors},
	urldate = {2026-06-07},
	volume = {612},
	year = 2022}

@article{wangQuantumStatesIntertwining2023,
	author = {Wang, Yaojia and Wu, Heng and McCandless, Gregory T. and Chan, Julia Y. and Ali, Mazhar N.},
	copyright = {2023 Springer Nature Limited},
	doi = {10.1038/s42254-023-00635-7},
	issn = {2522-5820},
	journal = {Nature Reviews Physics},
	langid = {english},
	month = nov,
	number = {11},
	pages = {635--658},
	publisher = {Nature Publishing Group},
	title = {Quantum States and Intertwining Phases in Kagome Materials},
	urldate = {2026-06-10},
	volume = {5},
	year = 2023}

@article{negiMagneticKagomeMaterials2025,
	author = {Negi, Pranav and Medhi, Koushik and Pancholi, Abhinav and Roychowdhury, Subhajit},
	doi = {10.1039/D5MH00120J},
	issn = {2051-6347, 2051-6355},
	journal = {Materials Horizons},
	langid = {english},
	number = {13},
	pages = {4510--4544},
	shorttitle = {Magnetic {{Kagome}} Materials},
	title = {Magnetic {{Kagome}} Materials: Bridging Fundamental Properties and Topological Quantum Applications},
	urldate = {2026-06-07},
	volume = {12},
	year = 2025}

@article{disanteKagomeMetals2026,
	author = {Di Sante, Domenico and Neupert, Titus and Sangiovanni, Giorgio and Thomale, Ronny and Comin, Riccardo and Checkelsky, Joseph G. and Zeljkovic, Ilija and Wilson, Stephen D.},
	doi = {10.1103/1g9n-wm38},
	journal = {Reviews of Modern Physics},
	month = feb,
	number = {1},
	pages = {015002},
	publisher = {American Physical Society},
	title = {Kagome Metals},
	urldate = {2026-05-10},
	volume = {98},
	year = 2026}

@article{wilsonAV3Sb5KagomeSuperconductors2024,
	author = {Wilson, Stephen D. and Ortiz, Brenden R.},
	copyright = {2024 Springer Nature Limited},
	doi = {10.1038/s41578-024-00677-y},
	issn = {2058-8437},
	journal = {Nature Reviews Materials},
	langid = {english},
	month = jun,
	number = {6},
	pages = {420--432},
	publisher = {Nature Publishing Group},
	title = {{{AV3Sb5}} Kagome Superconductors},
	urldate = {2026-06-10},
	volume = {9},
	year = 2024}

@article{ranProjectedWaveFunctionStudySpin2007,
	author = {Ran, Ying and Hermele, Michael and Lee, Patrick A. and Wen, Xiao-Gang},
	doi = {10.1103/PhysRevLett.98.117205},
	journal = {Physical Review Letters},
	month = mar,
	number = {11},
	pages = {117205},
	publisher = {American Physical Society},
	title = {Projected-Wave-Function Study of the Spin-$1/2$ Heisenberg Model on the Kagom\'e Lattice},
	urldate = {2026-06-12},
	volume = {98},
	year = 2007}

@article{balentsSpinLiquidsFrustrated2010,
	author = {Balents, Leon},
	copyright = {2010 Springer Nature Limited},
	doi = {10.1038/nature08917},
	issn = {1476-4687},
	journal = {Nature},
	langid = {english},
	month = mar,
	number = {7286},
	pages = {199--208},
	publisher = {Nature Publishing Group},
	title = {Spin Liquids in Frustrated Magnets},
	urldate = {2026-06-12},
	volume = {464},
	year = 2010}

@article{yanSpinLiquidGroundState2011,
	author = {Yan, Simeng and Huse, David A. and White, Steven R.},
	doi = {10.1126/science.1201080},
	journal = {Science},
	month = jun,
	number = {6034},
	pages = {1173--1176},
	publisher = {American Association for the Advancement of Science},
	title = {Spin-{{Liquid Ground State}} of the {{S}} = 1/2 {{Kagome Heisenberg Antiferromagnet}}},
	urldate = {2026-06-12},
	volume = {332},
	year = 2011}

@article{wangCompetingElectronicOrders2013,
	author = {Wang, Wan-Sheng and Li, Zheng-Zhao and Xiang, Yuan-Yuan and Wang, Qiang-Hua},
	doi = {10.1103/PhysRevB.87.115135},
	journal = {Physical Review B},
	month = mar,
	number = {11},
	pages = {115135},
	publisher = {American Physical Society},
	title = {Competing Electronic Orders on Kagome Lattices at van {{Hove}} Filling},
	urldate = {2026-06-12},
	volume = {87},
	year = 2013}

@article{xuIntrinsicQuantumAnomalous2015,
	author = {Xu, Gang and Lian, Biao and Zhang, Shou-Cheng},
	doi = {10.1103/PhysRevLett.115.186802},
	journal = {Physical Review Letters},
	month = oct,
	number = {18},
	pages = {186802},
	publisher = {American Physical Society},
	title = {Intrinsic Quantum Anomalous Hall Effect in the Kagome Lattice ${\mathrm{Cs}}_{2}{\mathrm{LiMn}}_{3}{\mathrm{F}}_{12}$},
	urldate = {2026-06-12},
	volume = {115},
	year = 2015}

@article{liRealizationFlatBand2018,
	author = {Li, Zhi and Zhuang, Jincheng and Wang, Li and Feng, Haifeng and Gao, Qian and Xu, Xun and Hao, Weichang and Wang, Xiaolin and Zhang, Chao and Wu, Kehui and Dou, Shi Xue and Chen, Lan and Hu, Zhenpeng and Du, Yi},
	doi = {10.1126/sciadv.aau4511},
	journal = {Science Advances},
	month = nov,
	number = {11},
	pages = {eaau4511},
	publisher = {American Association for the Advancement of Science},
	title = {Realization of Flat Band with Possible Nontrivial Topology in Electronic {{Kagome}} Lattice},
	urldate = {2026-06-12},
	volume = {4},
	year = 2018}

@article{grandiTheoryNematicCharge2023,
	author = {Grandi, Francesco and Consiglio, Armando and Sentef, Michael A. and Thomale, Ronny and Kennes, Dante M.},
	doi = {10.1103/PhysRevB.107.155131},
	journal = {Physical Review B},
	month = apr,
	number = {15},
	pages = {155131},
	publisher = {American Physical Society},
	title = {Theory of Nematic Charge Orders in Kagome Metals},
	urldate = {2026-06-12},
	volume = {107},
	year = 2023}

@article{yeMassiveDiracFermions2018,
	author = {Ye, Linda and Kang, Mingu and Liu, Junwei and {von Cube}, Felix and Wicker, Christina R. and Suzuki, Takehito and Jozwiak, Chris and Bostwick, Aaron and Rotenberg, Eli and Bell, David C. and Fu, Liang and Comin, Riccardo and Checkelsky, Joseph G.},
	copyright = {2018 Macmillan Publishers Limited, part of Springer Nature. All rights reserved.},
	doi = {10.1038/nature25987},
	issn = {1476-4687},
	journal = {Nature},
	langid = {english},
	month = mar,
	number = {7698},
	pages = {638--642},
	publisher = {Nature Publishing Group},
	title = {Massive {{Dirac}} Fermions in a Ferromagnetic Kagome Metal},
	urldate = {2026-06-08},
	volume = {555},
	year = 2018}

@article{kangDiracFermionsFlat2020,
	author = {Kang, Mingu and Ye, Linda and Fang, Shiang and You, Jhih-Shih and Levitan, Abe and Han, Minyong and Facio, Jorge I. and Jozwiak, Chris and Bostwick, Aaron and Rotenberg, Eli and Chan, Mun K. and McDonald, Ross D. and Graf, David and Kaznatcheev, Konstantine and Vescovo, Elio and Bell, David C. and Kaxiras, Efthimios and {van den Brink}, Jeroen and Richter, Manuel and Prasad Ghimire, Madhav and Checkelsky, Joseph G. and Comin, Riccardo},
	copyright = {2019 The Author(s), under exclusive licence to Springer Nature Limited},
	doi = {10.1038/s41563-019-0531-0},
	issn = {1476-4660},
	journal = {Nature Materials},
	langid = {english},
	month = feb,
	number = {2},
	pages = {163--169},
	publisher = {Nature Publishing Group},
	title = {Dirac Fermions and Flat Bands in the Ideal Kagome Metal {{FeSn}}},
	urldate = {2026-06-12},
	volume = {19},
	year = 2020}

@article{yeHoppingFrustrationinducedFlat2024,
	author = {Ye, Linda and Fang, Shiang and Kang, Mingu and Kaufmann, Josef and Lee, Yonghun and John, Caolan and Neves, Paul M. and Zhao, S. Y. Frank and Denlinger, Jonathan and Jozwiak, Chris and Bostwick, Aaron and Rotenberg, Eli and Kaxiras, Efthimios and Bell, David C. and Janson, Oleg and Comin, Riccardo and Checkelsky, Joseph G.},
	copyright = {2024 The Author(s), under exclusive licence to Springer Nature Limited},
	doi = {10.1038/s41567-023-02360-5},
	issn = {1745-2481},
	journal = {Nature Physics},
	langid = {english},
	month = apr,
	number = {4},
	pages = {610--614},
	publisher = {Nature Publishing Group},
	title = {Hopping Frustration-Induced Flat Band and Strange Metallicity in a Kagome Metal},
	urldate = {2026-06-12},
	volume = {20},
	year = 2024}

@article{ortizNewKagomePrototype2019,
	author = {Ortiz, Brenden R. and Gomes, L{\'\i}dia C. and Morey, Jennifer R. and Winiarski, Michal and Bordelon, Mitchell and Mangum, John S. and Oswald, Iain W. H. and {Rodriguez-Rivera}, Jose A. and Neilson, James R. and Wilson, Stephen D. and Ertekin, Elif and McQueen, Tyrel M. and Toberer, Eric S.},
	doi = {10.1103/PhysRevMaterials.3.094407},
	journal = {Physical Review Materials},
	month = sep,
	number = {9},
	pages = {094407},
	publisher = {American Physical Society},
	shorttitle = {New Kagome Prototype Materials},
	title = {New kagome prototype materials: discovery of ${\mathrm{KV}}_{3}{\mathrm{Sb}}_{5},{\mathrm{RbV}}_{3}{\mathrm{Sb}}_{5}$, and ${\mathrm{CsV}}_{3}{\mathrm{Sb}}_{5}$},
	urldate = {2026-06-10},
	volume = {3},
	year = 2019}

@article{ortizCs135_Z2Topological2020,
	author = {Ortiz, Brenden R. and Teicher, Samuel M. L. and Hu, Yong and Zuo, Julia L. and Sarte, Paul M. and Schueller, Emily C. and Abeykoon, A. M. Milinda and Krogstad, Matthew J. and Rosenkranz, Stephan and Osborn, Raymond and Seshadri, Ram and Balents, Leon and He, Junfeng and Wilson, Stephen D.},
	doi = {10.1103/PhysRevLett.125.247002},
	journal = {Physical Review Letters},
	month = dec,
	number = {24},
	pages = {247002},
	publisher = {American Physical Society},
	shorttitle = {\$\textbackslash mathrm\textbraceleft{{Cs}}\textbraceright\textbraceleft\textbackslash mathrm\textbraceleft{{V}}\textbraceright\textbraceright\_\textbraceleft 3\textbraceright\textbraceleft\textbackslash mathrm\textbraceleft{{Sb}}\textbraceright\textbraceright\_\textbraceleft 5\textbraceright\$},
	title = {$\mathrm{Cs}{\mathrm{V}}_{3}{\mathrm{Sb}}_{5}$: A ${\mathbb{Z}}_{2}$ Topological Kagome Metal with a Superconducting Ground State},
	urldate = {2026-06-12},
	volume = {125},
	year = 2020}

@article{ortizSuperconductivityZ2Kagome2021,
	author = {Ortiz, Brenden R. and Sarte, Paul M. and Kenney, Eric M. and Graf, Michael J. and Teicher, Samuel M. L. and Seshadri, Ram and Wilson, Stephen D.},
	doi = {10.1103/PhysRevMaterials.5.034801},
	journal = {Physical Review Materials},
	month = mar,
	number = {3},
	pages = {034801},
	publisher = {American Physical Society},
	title = {Superconductivity in the ${\mathbb{Z}}_{2}$ kagome metal ${\mathrm{KV}}_{3}{\mathrm{Sb}}_{5}$},
	urldate = {2026-06-12},
	volume = {5},
	year = 2021}

@article{niAnisotropicSuperconductingProperties2021,
	author = {Ni, Shunli and Ma, Sheng and Zhang, Yuhang and Yuan, Jie and Yang, Haitao and Lu, Zouyouwei and Wang, Ningning and Sun, Jianping and Zhao, Zhen and Li, Dong and Liu, Shaobo and Zhang, Hua and Chen, Hui and Jin, Kui and Cheng, Jinguang and Yu, Li and Zhou, Fang and Dong, Xiaoli and Hu, Jiangping and Gao, Hong-Jun and Zhao, Zhongxian},
	copyright = {http://creativecommons.org/licenses/by/3.0/},
	doi = {10.1088/0256-307X/38/5/057403},
	issn = {0256-307X},
	journal = {Chin. Phys. Lett.},
	langid = {english},
	month = may,
	number = {5},
	pages = {057403--057403},
	title = {Anisotropic {{Superconducting Properties}} of {{Kagome Metal CsV}}\_{{3Sb}}\_5},
	urldate = {2026-06-12},
	volume = {38},
	year = 2021}

@article{yinSuperconductivityNormalStateProperties2021,
	author = {Yin, Qiangwei and Tu, Zhijun and Gong, Chunsheng and Fu, Yang and Yan, Shaohua and Lei, Hechang},
	copyright = {http://creativecommons.org/licenses/by/3.0/},
	doi = {10.1088/0256-307X/38/3/037403},
	issn = {0256-307X},
	journal = {Chin. Phys. Lett.},
	langid = {english},
	month = mar,
	number = {3},
	pages = {037403--037403},
	title = {Superconductivity and {{Normal-State Properties}} of {{Kagome Metal RbV}}\_{{3Sb}}\_5 {{Single Crystals}}},
	urldate = {2026-06-12},
	volume = {38},
	year = 2021}

@article{duanNodelessSuperconductivityKagome2021,
	author = {Duan, Weiyin and Nie, Zhiyong and Luo, Shuaishuai and Yu, Fanghang and Ortiz, Brenden R. and Yin, Lichang and Su, Hang and Du, Feng and Wang, An and Chen, Ye and Lu, Xin and Ying, Jianjun and Wilson, Stephen D. and Chen, Xianhui and Song, Yu and Yuan, Huiqiu},
	doi = {10.1007/s11433-021-1747-7},
	issn = {1869-1927},
	journal = {Science China Physics, Mechanics \& Astronomy},
	langid = {english},
	month = jul,
	number = {10},
	pages = {107462},
	title = {Nodeless Superconductivity in the Kagome Metal {{CsV3Sb5}}},
	urldate = {2026-06-12},
	volume = {64},
	year = 2021}

@article{wangCompetitionChargedensitywaveSuperconductivity2021,
	author = {Wang, N. N. and Chen, K. Y. and Yin, Q. W. and Ma, Y. N. N. and Pan, B. Y. and Yang, X. and Ji, X. Y. and Wu, S. L. and Shan, P. F. and Xu, S. X. and Tu, Z. J. and Gong, C. S. and Liu, G. T. and Li, G. and Uwatoko, Y. and Dong, X. L. and Lei, H. C. and Sun, J. P. and Cheng, J.-G.},
	doi = {10.1103/PhysRevResearch.3.043018},
	journal = {Physical Review Research},
	month = oct,
	number = {4},
	pages = {043018},
	publisher = {American Physical Society},
	title = {Competition between charge-density-wave and superconductivity in the kagome metal $\mathrm{Rb}{\mathrm{V}}_{3}{\mathrm{Sb}}_{5}$},
	urldate = {2026-06-12},
	volume = {3},
	year = 2021}

@article{chenDoubleSuperconductingDome2021,
	author = {Chen, K. Y. and Wang, N. N. and Yin, Q. W. and Gu, Y. H. and Jiang, K. and Tu, Z. J. and Gong, C. S. and Uwatoko, Y. and Sun, J. P. and Lei, H. C. and Hu, J. P. and Cheng, J.-G.},
	doi = {10.1103/PhysRevLett.126.247001},
	journal = {Physical Review Letters},
	month = jun,
	number = {24},
	pages = {247001},
	publisher = {American Physical Society},
	title = {Double Superconducting Dome and Triple Enhancement of ${T}_{c}$ in the Kagome Superconductor ${\mathrm{CsV}}_{3}{\mathrm{Sb}}_{5}$ under High Pressure},
	urldate = {2026-06-12},
	volume = {126},
	year = 2021}

@article{tanChargeDensityWaves2021,
	author = {Tan, Hengxin and Liu, Yizhou and Wang, Ziqiang and Yan, Binghai},
	doi = {10.1103/PhysRevLett.127.046401},
	journal = {Physical Review Letters},
	month = jul,
	number = {4},
	pages = {046401},
	publisher = {American Physical Society},
	title = {Charge {{Density Waves}} and {{Electronic Properties}} of {{Superconducting Kagome Metals}}},
	urldate = {2026-06-12},
	volume = {127},
	year = 2021}

@article{zhaoCascadeCorrelatedElectron2021,
	author = {Zhao, He and Li, Hong and Ortiz, Brenden R. and Teicher, Samuel M. L. and Park, Takamori and Ye, Mengxing and Wang, Ziqiang and Balents, Leon and Wilson, Stephen D. and Zeljkovic, Ilija},
	copyright = {2021 The Author(s), under exclusive licence to Springer Nature Limited},
	doi = {10.1038/s41586-021-03946-w},
	issn = {1476-4687},
	journal = {Nature},
	langid = {english},
	month = nov,
	number = {7884},
	pages = {216--221},
	publisher = {Nature Publishing Group},
	title = {Cascade of Correlated Electron States in the Kagome Superconductor {{CsV3Sb5}}},
	urldate = {2026-06-12},
	volume = {599},
	year = 2021}

@article{chenRotonPairDensity2021,
	author = {Chen, Hui and Yang, Haitao and Hu, Bin and Zhao, Zhen and Yuan, Jie and Xing, Yuqing and Qian, Guojian and Huang, Zihao and Li, Geng and Ye, Yuhan and Ma, Sheng and Ni, Shunli and Zhang, Hua and Yin, Qiangwei and Gong, Chunsheng and Tu, Zhijun and Lei, Hechang and Tan, Hengxin and Zhou, Sen and Shen, Chengmin and Dong, Xiaoli and Yan, Binghai and Wang, Ziqiang and Gao, Hong-Jun},
	copyright = {2021 The Author(s), under exclusive licence to Springer Nature Limited},
	doi = {10.1038/s41586-021-03983-5},
	issn = {1476-4687},
	journal = {Nature},
	langid = {english},
	month = nov,
	number = {7884},
	pages = {222--228},
	publisher = {Nature Publishing Group},
	title = {Roton Pair Density Wave in a Strong-Coupling Kagome Superconductor},
	urldate = {2026-06-12},
	volume = {599},
	year = 2021}

@article{mielkeTimereversalSymmetrybreakingCharge2022,
	author = {Mielke, C. and Das, D. and Yin, J.-X. and Liu, H. and Gupta, R. and Jiang, Y.-X. and Medarde, M. and Wu, X. and Lei, H. C. and Chang, J. and Dai, Pengcheng and Si, Q. and Miao, H. and Thomale, R. and Neupert, T. and Shi, Y. and Khasanov, R. and Hasan, M. Z. and Luetkens, H. and Guguchia, Z.},
	copyright = {2022 The Author(s), under exclusive licence to Springer Nature Limited},
	doi = {10.1038/s41586-021-04327-z},
	issn = {1476-4687},
	journal = {Nature},
	langid = {english},
	month = feb,
	number = {7896},
	pages = {245--250},
	publisher = {Nature Publishing Group},
	title = {Time-Reversal Symmetry-Breaking Charge Order in a Kagome Superconductor},
	urldate = {2026-06-12},
	volume = {602},
	year = 2022}

@article{luoElectronicNatureCharge2022,
	author = {Luo, Hailan and Gao, Qiang and Liu, Hongxiong and Gu, Yuhao and Wu, Dingsong and Yi, Changjiang and Jia, Junjie and Wu, Shilong and Luo, Xiangyu and Xu, Yu and Zhao, Lin and Wang, Qingyan and Mao, Hanqing and Liu, Guodong and Zhu, Zhihai and Shi, Youguo and Jiang, Kun and Hu, Jiangping and Xu, Zuyan and Zhou, X. J.},
	copyright = {2022 The Author(s)},
	doi = {10.1038/s41467-021-27946-6},
	issn = {2041-1723},
	journal = {Nature Communications},
	langid = {english},
	month = jan,
	number = {1},
	pages = {273},
	publisher = {Nature Publishing Group},
	title = {Electronic Nature of Charge Density Wave and Electron-Phonon Coupling in Kagome Superconductor {{KV3Sb5}}},
	urldate = {2026-06-10},
	volume = {13},
	year = 2022}

@article{kangTwofoldVanHove2022,
	author = {Kang, Mingu and Fang, Shiang and Kim, Jeong-Kyu and Ortiz, Brenden R. and Ryu, Sae Hee and Kim, Jimin and Yoo, Jonggyu and Sangiovanni, Giorgio and Di Sante, Domenico and Park, Byeong-Gyu and Jozwiak, Chris and Bostwick, Aaron and Rotenberg, Eli and Kaxiras, Efthimios and Wilson, Stephen D. and Park, Jae-Hoon and Comin, Riccardo},
	copyright = {2022 The Author(s), under exclusive licence to Springer Nature Limited},
	doi = {10.1038/s41567-021-01451-5},
	issn = {1745-2481},
	journal = {Nature Physics},
	langid = {english},
	month = mar,
	number = {3},
	pages = {301--308},
	publisher = {Nature Publishing Group},
	title = {Twofold van {{Hove}} Singularity and Origin of Charge Order in Topological Kagome Superconductor {{CsV3Sb5}}},
	urldate = {2026-06-12},
	volume = {18},
	year = 2022}

@article{huRichNatureVan2022,
	author = {Hu, Yong and Wu, Xianxin and Ortiz, Brenden R. and Ju, Sailong and Han, Xinloong and Ma, Junzhang and Plumb, Nicholas C. and Radovic, Milan and Thomale, Ronny and Wilson, Stephen D. and Schnyder, Andreas P. and Shi, Ming},
	copyright = {2022 The Author(s)},
	doi = {10.1038/s41467-022-29828-x},
	issn = {2041-1723},
	journal = {Nature Communications},
	langid = {english},
	month = apr,
	number = {1},
	pages = {2220},
	publisher = {Nature Publishing Group},
	title = {Rich Nature of {{Van Hove}} Singularities in {{Kagome}} Superconductor {{CsV3Sb5}}},
	urldate = {2026-06-12},
	volume = {13},
	year = 2022}

@article{guoSwitchableChiralTransport2022,
	author = {Guo, Chunyu and Putzke, Carsten and Konyzheva, Sofia and Huang, Xiangwei and {Gutierrez-Amigo}, Martin and Errea, Ion and Chen, Dong and Vergniory, Maia G. and Felser, Claudia and Fischer, Mark H. and Neupert, Titus and Moll, Philip J. W.},
	copyright = {2022 The Author(s)},
	doi = {10.1038/s41586-022-05127-9},
	issn = {1476-4687},
	journal = {Nature},
	langid = {english},
	month = nov,
	number = {7936},
	pages = {461--466},
	publisher = {Nature Publishing Group},
	title = {Switchable Chiral Transport in Charge-Ordered Kagome Metal {{CsV3Sb5}}},
	urldate = {2026-06-12},
	volume = {611},
	year = 2022}

@article{neupertChargeOrderSuperconductivity2022,
	author = {Neupert, Titus and Denner, M. Michael and Yin, Jia-Xin and Thomale, Ronny and Hasan, M. Zahid},
	copyright = {2021 Springer Nature Limited},
	doi = {10.1038/s41567-021-01404-y},
	issn = {1745-2481},
	journal = {Nature Physics},
	langid = {english},
	month = feb,
	number = {2},
	pages = {137--143},
	publisher = {Nature Publishing Group},
	title = {Charge Order and Superconductivity in Kagome Materials},
	urldate = {2026-06-10},
	volume = {18},
	year = 2022}

@article{zhengEmergentChargeOrder2022,
	author = {Zheng, Lixuan and Wu, Zhimian and Yang, Ye and Nie, Linpeng and Shan, Min and Sun, Kuanglv and Song, Dianwu and Yu, Fanghang and Li, Jian and Zhao, Dan and Li, Shunjiao and Kang, Baolei and Zhou, Yanbing and Liu, Kai and Xiang, Ziji and Ying, Jianjun and Wang, Zhenyu and Wu, Tao and Chen, Xianhui},
	copyright = {2022 The Author(s), under exclusive licence to Springer Nature Limited},
	doi = {10.1038/s41586-022-05351-3},
	issn = {1476-4687},
	journal = {Nature},
	langid = {english},
	month = nov,
	number = {7937},
	pages = {682--687},
	publisher = {Nature Publishing Group},
	title = {Emergent Charge Order in Pressurized Kagome Superconductor {{CsV3Sb5}}},
	urldate = {2026-06-12},
	volume = {611},
	year = 2022}

@article{nieChargedensitywavedrivenElectronicNematicity2022,
	author = {Nie, Linpeng and Sun, Kuanglv and Ma, Wanru and Song, Dianwu and Zheng, Lixuan and Liang, Zuowei and Wu, Ping and Yu, Fanghang and Li, Jian and Shan, Min and Zhao, Dan and Li, Shunjiao and Kang, Baolei and Wu, Zhimian and Zhou, Yanbing and Liu, Kai and Xiang, Ziji and Ying, Jianjun and Wang, Zhenyu and Wu, Tao and Chen, Xianhui},
	copyright = {2022 The Author(s), under exclusive licence to Springer Nature Limited},
	doi = {10.1038/s41586-022-04493-8},
	issn = {1476-4687},
	journal = {Nature},
	langid = {english},
	month = apr,
	number = {7904},
	pages = {59--64},
	publisher = {Nature Publishing Group},
	title = {Charge-Density-Wave-Driven Electronic Nematicity in a Kagome Superconductor},
	urldate = {2026-06-12},
	volume = {604},
	year = 2022}

@article{guoManybodyInterferenceKagome2025,
	author = {Guo, Chunyu and Wang, Kaize and Zhang, Ling and Putzke, Carsten and Chen, Dong and {van Delft}, Maarten R. and Wiedmann, Steffen and Balakirev, Fedor F. and McDonald, Ross D. and {Gutierrez-Amigo}, Martin and Alkorta, Manex and Errea, Ion and Vergniory, Maia G. and Oka, Takashi and Moessner, Roderich and Fischer, Mark H. and Neupert, Titus and Felser, Claudia and Moll, Philip J. W.},
	copyright = {2025 The Author(s)},
	doi = {10.1038/s41586-025-09659-8},
	issn = {1476-4687},
	journal = {Nature},
	langid = {english},
	month = nov,
	number = {8088},
	pages = {68--73},
	publisher = {Nature Publishing Group},
	title = {Many-Body Interference in Kagome Crystals},
	urldate = {2026-06-12},
	volume = {647},
	year = 2025}

@article{wangSoftPhononChargeDensityWave2026,
	author = {Wang, Yifan and Xu, Chenchao and Wu, Zhimian and Rao, Huachen and Shan, Zhaoyang and Liu, Yi and Cao, Guanghan and Smidman, Michael and Shi, Ming and Yuan, Huiqiu and Wu, Tao and Chen, Xianhui and Cao, Chao and Song, Yu},
	doi = {10.1103/lhfz-7tk1},
	journal = {Physical Review Letters},
	month = mar,
	number = {13},
	pages = {136401},
	publisher = {American Physical Society},
	title = {Soft Phonon Charge-Density-Wave Formation in the Kagome Metal ${\mathrm{KV}}_{3}{\mathrm{Sb}}_{5}$},
	urldate = {2026-06-10},
	volume = {136},
	year = 2026}

@article{liuSuperconductivityPressureChromiumbased2024,
	author = {Liu, Yi and Liu, Zi-Yi and Bao, Jin-Ke and Yang, Peng-Tao and Ji, Liang-Wen and Wu, Si-Qi and Shen, Qin-Xin and Luo, Jun and Yang, Jie and Liu, Ji-Yong and Xu, Chen-Chao and Yang, Wu-Zhang and Chai, Wan-Li and Lu, Jia-Yi and Liu, Chang-Chao and Wang, Bo-Sen and Jiang, Hao and Tao, Qian and Ren, Zhi and Xu, Xiao-Feng and Cao, Chao and Xu, Zhu-An and Zhou, Rui and Cheng, Jin-Guang and Cao, Guang-Han},
	copyright = {2024 The Author(s), under exclusive licence to Springer Nature Limited},
	doi = {10.1038/s41586-024-07761-x},
	issn = {1476-4687},
	journal = {Nature},
	langid = {english},
	month = aug,
	number = {8027},
	pages = {1032--1037},
	publisher = {Nature Publishing Group},
	title = {Superconductivity under Pressure in a Chromium-Based Kagome Metal},
	urldate = {2026-06-12},
	volume = {632},
	year = 2024}

@article{pengFlatBandsDistinct2026,
	archiveprefix = {arXiv},
	author = {Peng, Shuting and Han, Yulei and Li, Yongkai and Shen, Jianchang and Miao, Yu and Luo, Yang and Huai, Linwei and Ou, Zhipeng and Li, Hongyu and Xiang, Ziji and Liu, Zhengtai and Shen, Dawei and Hashimoto, Makoto and Lu, Donghui and Yao, Yugui and Qiao, Zhenhua and Wang, Zhiwei and He, Junfeng},
	doi = {10.1007/s11433-025-2773-3},
	eprint = {2406.17769},
	issn = {1674-7348, 1869-1927},
	journal = {Science China Physics, Mechanics \& Astronomy},
	month = jan,
	number = {1},
	pages = {217412},
	primaryclass = {cond-mat.supr-con},
	title = {Flat Bands and Distinct Density Wave Orders in Correlated {{Kagome}} Superconductor CsCr3Sb5},
	urldate = {2026-06-12},
	volume = {69},
	year = 2026}

@article{liElectronCorrelationIncipient2025,
	author = {Li, Yidian and Liu, Yi and Du, Xian and Wu, Siqi and Zhao, Wenxuan and Zhai, Kaiyi and Hu, Yinqi and Zhang, Senyao and Chen, Houke and Liu, Jieyi and Yang, Yiheng and Peng, Cheng and Hashimoto, Makoto and Lu, Donghui and Liu, Zhongkai and Wang, Yilin and Chen, Yulin and Cao, Guanghan and Yang, Lexian},
	copyright = {2025 The Author(s)},
	doi = {10.1038/s41467-025-58487-x},
	issn = {2041-1723},
	journal = {Nature Communications},
	langid = {english},
	month = apr,
	number = {1},
	pages = {3229},
	publisher = {Nature Publishing Group},
	title = {Electron Correlation and Incipient Flat Bands in the {{Kagome}} Superconductor {{CsCr3Sb5}}},
	urldate = {2026-06-12},
	volume = {16},
	year = 2025}

@article{wangSpinExcitationsFlat2025,
	author = {Wang, Zehao and Guo, Yucheng and Huang, Hsiao-Yu and Xie, Fang and Huang, Yuefei and Gao, Bin and Oh, Ji Seop and Wu, Han and Okamoto, Jun and Channagowdra, Ganesha and Chen, Chien-Te and Ye, Feng and Lu, Xingye and Liu, Zhaoyu and Ren, Zheng and Fang, Yuan and Wang, Yiming and Biswas, Ananya and Zhang, Yichen and Yue, Ziqin and Hu, Cheng and Jozwiak, Chris and Bostwick, Aaron and Rotenberg, Eli and Hashimoto, Makoto and Lu, Donghui and Kono, Junichiro and Chu, Jiun-Haw and Yakobson, Boris I. and Birgeneau, Robert J. and Cao, Guang-Han and Fujimori, Atsushi and Huang, Di-Jing and Si, Qimiao and Yi, Ming and Dai, Pengcheng},
	copyright = {2025 The Author(s)},
	doi = {10.1038/s41467-025-62298-5},
	issn = {2041-1723},
	journal = {Nature Communications},
	langid = {english},
	month = aug,
	number = {1},
	pages = {7573},
	publisher = {Nature Publishing Group},
	title = {Spin Excitations and Flat Electronic Bands in a {{Cr-based}} Kagome Superconductor},
	urldate = {2026-06-12},
	volume = {16},
	year = 2025}

@article{xuAltermagneticGroundState2025,
	author = {Xu, Chenchao and Wu, Siqi and Zhi, Guo-Xiang and Cao, Guanghan and Dai, Jianhui and Cao, Chao and Wang, Xiaoqun and Lin, Hai-Qing},
	copyright = {2025 The Author(s)},
	doi = {10.1038/s41467-025-58446-6},
	issn = {2041-1723},
	journal = {Nature Communications},
	langid = {english},
	month = apr,
	number = {1},
	pages = {3114},
	publisher = {Nature Publishing Group},
	title = {Altermagnetic Ground State in Distorted {{Kagome}} Metal {{CsCr3Sb5}}},
	urldate = {2026-06-10},
	volume = {16},
	year = 2025}

@article{xieElectronCorrelationsKagome2025,
	author = {Xie, Fang and Fang, Yuan and Li, Ying and Huang, Yuefei and Chen, Lei and Setty, Chandan and Sur, Shouvik and Yakobson, Boris and Valent{\'\i}, Roser and Si, Qimiao},
	doi = {10.1103/PhysRevResearch.7.L022061},
	journal = {Physical Review Research},
	month = jun,
	number = {2},
	pages = {L022061},
	publisher = {American Physical Society},
	title = {Electron correlations in the kagome flat band metal ${\mathrm{CsCr}}_{3}{\mathrm{Sb}}_{5}$},
	urldate = {2026-06-12},
	volume = {7},
	year = 2025}

@article{liuUnravelingIntertwinedOrders2026,
	author = {Liu, Liangyang and Li, Yidian and Tan, Hengxin and Liu, Yi and Sun, Kuanglv and Shi, Ying and Zhai, Yuxin and Lin, Hao and Cao, Guanghan and Yan, Binghai and Chen, Xianhui and Wu, Tao and Zhang, Guang-Ming and Yang, Luyi},
	doi = {10.1093/nsr/nwag044},
	issn = {2095-5138},
	journal = {National Science Review},
	month = mar,
	number = {6},
	pages = {nwag044},
	title = {Unraveling Intertwined Orders in the Strongly Correlated Kagome Metal {{CsCr3Sb5}}},
	urldate = {2026-06-12},
	volume = {13},
	year = 2026}

@article{chengFriezeChargeStripes2026,
	author = {Cheng, Siyu and Zeng, Keyu and Liu, Yi and Candelora, Christopher and Wang, Ziqiang and Cao, Guang-Han and Zeljkovic, Ilija},
	copyright = {2026 The Author(s), under exclusive licence to Springer Nature Limited},
	doi = {10.1038/s41567-026-03232-4},
	issn = {1745-2481},
	journal = {Nature Physics},
	langid = {english},
	month = may,
	number = {5},
	pages = {706--712},
	publisher = {Nature Publishing Group},
	title = {Frieze Charge Stripes in a Correlated Kagome Superconductor},
	urldate = {2026-06-12},
	volume = {22},
	year = 2026}

@article{wuFlatbandEnhancedAntiferromagnetic2025a,
	author = {Wu, Siqi and Xu, Chenchao and Wang, Xiaoqun and Lin, Hai-Qing and Cao, Chao and Cao, Guang-Han},
	copyright = {2025 The Author(s)},
	doi = {10.1038/s41467-025-56582-7},
	issn = {2041-1723},
	journal = {Nature Communications},
	langid = {english},
	month = feb,
	number = {1},
	pages = {1375},
	publisher = {Nature Publishing Group},
	title = {Flat-Band Enhanced Antiferromagnetic Fluctuations and Superconductivity in Pressurized {{CsCr3Sb5}}},
	urldate = {2026-06-10},
	volume = {16},
	year = 2025}

\end{document}